\documentclass[referee]{raa}

\usepackage{graphicx,times}
\usepackage{natbib}
\usepackage{amssymb,amsmath}
\bibpunct{(}{)}{;}{a}{}{,}

\usepackage[pagebackref=true]{hyperref}
\usepackage{threeparttable}

\begin{document}

  \title{Formation of Extremely Low-Mass White Dwarfs in Wide Orbits}

   \volnopage{Vol.0 (202x) No.0, 000--000}      
   \setcounter{page}{1}          

   \author{Yangyang Zhang 
      \inst{1, *} \footnotetext{$*$E-mail:zhangyy@zknu.edu.cn}
   \and Zhenwei Li
      \inst{2, 3, 4, *}\footnotetext{$*$E-mail:lizw@ynao.ac.cn}
   \and Xuefei Chen
      \inst{2, 3, 4, *}\footnotetext{$*$E-mail:cxf@ynao.ac.cn}
   }

   \institute{Zhoukou Normal University, Zhoukou 466001+, China; {\it zhangyy@zknu.edu.cn}\\
        \and
             Yunnan Observatories, Chinese Academy of Sciences, Kunming 650216, China; {\it lizw@ynao.ac.cn, cxf@ynao.ac.cn}\\
        \and
             Key Laboratory for the Structure and Evolution of Celestial Objects, Chinese Academy of Sciences, Kunming 650216, China;\\
        \and
             Center for Astronomical Mega-Science, Chinese Academy of Sciences, 20A Datun Road, Chaoyang District, Beijing, 100012, China\\
\vs\no
   {\small Received 202x month day; accepted 202x month day}}

\abstract{Helium white dwarfs (WDs) with masses less than 0.3 $\rm M_{\sun}$ are known as extremely low-mass WDs (ELM WDs), which cannot be produced by single stellar evolution in theory. Generally, these stars are believed to form through binary interactions. Recently, two ELM WDs in unusually wide orbits were reported, i.e., KIC 8145411 and HE 0430-2457. Their orbital separations are too wide to be produced by the binary evolution scenario. In this work, we study the formation of wide-orbit ELM WD binaries from hierarchical triple systems. In this scenario, an ELM WD is formed from the inner binary and subsequently forms a wide binary system with the third object. We find that the merger of an evolved star with a brown dwarf in the inner binary fails to produce single ELM WDs, but Type Ia supernovae (SNe Ia) explosions can successfully do so. Furthermore, we investigate the impact of the supernova explosion on the orbital distribution of the surviving binary and find that this channel may have a probability of reproducing the orbital parameters of HE 0430-2457, but fails to reproduce the observed features of KIC 8145411. This supports recent observational recalibrations suggesting that KIC 8145411 resides in a triple system rather than a binary.
}
   \authorrunning{Y.Zhang et al.}            
   \titlerunning{ELM WDs in wide orbit }  
   \maketitle
%
%

\section{Introduction}           
\label{sec:1}
  
Helium white dwarfs (WDs) with masses less than $\sim$0.30 $\rm M_{\sun}$ are known as extremely low-mass WDs (ELM WDs), which have been largely found in the ELM Survey \citep{Brown2010, Brown2012, Brown2013, Brown2016a, Brown2020, Brown2022, Kilic2011a, Kilic2012, Gianninas2015}. 
To produce an ELM WD, the hydrogen (H)-rich envelope must be stripped before the star develops a massive helium core.
In theory, the Galaxy is not old enough to produce ELM WDs through single stellar evolution \citep{Iben1993}. 
Binary evolution is thought to be the main formation channel for ELM WDs. In this scenario, the envelope can be stripped by stable Roche-lobe overflow (RLOF) or common envelope (CE) ejection \citep{Marsh1995, Istrate2016, Chen2017, Sun2018, Li2019, Li2023}. Therefore, ELM WDs provide critical constraints on binary interaction physics. Moreover, many ELM WD binaries are found with very short orbital periods, and some are expected to be detectable by future space-borne gravitational wave antennas, e.g., LISA and Tianqin \citep{Brown2020, Li2020, LiuZhengwei2023, Amaro-Seoane2023}. 
For example, the most compact detached ELM WD binary, ZTF J1539+5027, with an orbital period of 6.91 minutes, has been confirmed as a gravitational wave source \citep{Burdge2020a}.

For the observed ELM WDs, two have been detected in extremely wide orbits, i.e., HE 0430-2457 with a semimajor axis of $\sim$1.59 au \citep{Vos2018b} and KIC 8145411 with a semimajor axis of $\sim$1.28 au \citep{Masuda2019} \footnote{The latest observations by \citet{Yamaguchi2024} reveal that this system is in fact a hierarchical triple containing a white dwarf with a mass of 0.53 $\rm M_{\sun}$, while the study by \citet{Masuda2019} underestimated the WD mass (0.2 $\rm M_{\sun}$) due to neglecting the light from the tertiary. Nevertheless, in this study we continue to treat KIC 8145411 as a wide-orbit binary system hosting an ELM WD (0.2 $\rm M_{\sun}$).}. 
We summarize their binary parameters in Table 1. 
The traditional binary evolution scenario has difficulty explaining the formation of these systems, as the progenitor radius of an ELM WD would be approximately ten times smaller than its Roche lobe radius \citep{Masuda2019}. \citet{Khurana2023} explained these systems' formation through binary-binary interactions. However, they also note that the probability of forming wide ELM WDs through these interactions is small.

\begin{figure}
  \centering
  \includegraphics[width=0.8\linewidth]{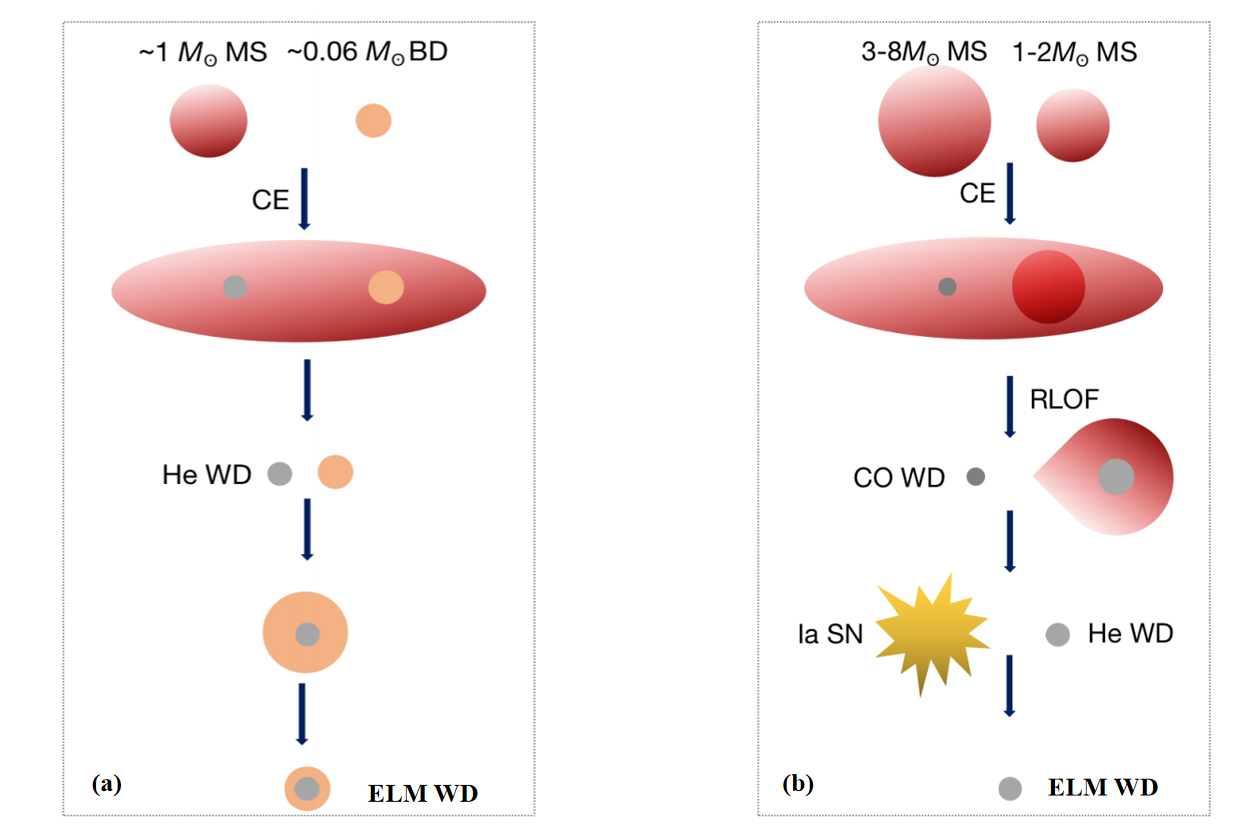} 
  \caption{Possible formation channels for single ELM WDs. Left panel: Merger of an evolved star with a brown dwarf. Right panel: SN explosion channel.}
  \label{Fig1}
\end{figure}

\citet{Vos2018b} proposed that hierarchical triple systems may contribute to the formation of wide-orbit ELM WD binaries, where the ELM WD is produced from the evolution of the inner binary, while the outer star does not directly interact with the inner binary. Thus, the problem can be simplified to the formation of single ELM WDs. One possible formation channel for single ELM WDs is the ejection of the WD from a binary system following a Type Ia supernova (SN Ia) of its companion \citep{Tauris1998, Justham2009, Meng2009, Wang2010}.
Additionally, another possible scenario is the merger of an evolved star with a brown dwarf (BD) \citep{Nelemans1998, Shao2012}. In this case, the donor's envelope can be stripped through spiral-in processes, after which the brown dwarf either evaporates or fills its Roche lobe, ultimately producing a single low-mass He WD. These two channels are described in Figure~\ref{Fig1}.

However, the triple systems that produce observed wide-orbit ELM WD binaries remain poorly constrained. In this paper, we consider hierarchical triple systems that contain an inner binary and a tertiary component, with the tertiary assumed to be an F/K-type star based on observations. We calculate the production of ELM WDs in triple systems through the above two channels, (i) the merger of an evolved star with a BD, and (ii) the SN explosion channel, thereby generating white dwarf binary systems with wide orbits, respectively.

This paper is organized as follows. In Section 2, we test whether the single ELM WD can be produced from the merger of an evolved star with a BD. In Section 3, we employ stellar evolution codes to simulate the formation of a single ELM WD from the SN explosion channel and characterize the distributions of orbital separation and eccentricity for the white dwarf-tertiary binary system. Then, we compare our results with the observational data. Section 4 provides a comprehensive summary and discussion of our work.

%
\begin{table}
\begin{center}
\begin{threeparttable}
\caption[]{The parameters of two observed binaries}\label{Tab1}


 \begin{tabular}{clcl}
  \hline\noalign{\smallskip}
                          &  KIC 8145411 \tnote{1}       & HE 0430-2457 \tnote{2}                  \\
  \hline\noalign{\smallskip}
$M_{\rm {ELM}} (\rm M_{\sun})$  & $0.20 \pm 0.009$     & $0.23 \pm 0.05$    \\ 
$M_{\rm {com}} (\rm M_{\sun})$  & $1.1 \pm 0.08$     &   $0.71 \pm 0.09$      \\
Period(days)  & $448.6_{-5.2}^{+5.8}$     &  $771 \pm 3$     \\
semi-major axis ($\rm R_{\sun}$)   & $274 \pm 6$ & $347_{-18}^{+15}$ \\
eccentricity  &  $0.14 \pm 0.02$  &  $\sim 0$\\
  \noalign{\smallskip}\hline
\end{tabular}
\begin{tablenotes}
       \footnotesize
        \item Note. ${}^{1}$ \citet{Masuda2019}; ${}^{2}$ \citet{Vos2018b}  
\end{tablenotes}
\end{threeparttable}
\end{center}
\end{table}

\section{Merger of an evolved star with a brown dwarf}
\label{sec:2}
We first investigate whether the inner binary can produce a single ELM WD through the merger of an evolved star with a brown dwarf, as shown in panel (a) of Figure~\ref{Fig1}. In this channel, the inner binary first evolves into a binary system consisting of an ELM WD and a brown dwarf through a common envelope phase. During the inner binary evolution, the lost material may interact with the tertiary star, thereby modifying the outer orbital separation \citep{Toonen2016}. For simplicity, we neglect the interactions between the tertiary component and the inner binary, and this assumption is also applied in the SN explosion channel. 

Brown dwarfs are degenerate objects that cannot fuse hydrogen in their cores. Several brown dwarfs with sdB and WD companions have been found in close orbits \citep{Casewell2020}, and they are supposed to be formed through the common envelope ejection process. The companion masses of brown dwarfs in observations are generally larger than 0.4 $\rm M_{\sun}$, and there are no observed samples of brown dwarfs with ELM WD companions \citep{Maxted2006, Beuermann2013, Parsons2017b, Zorotovic2022, Casewell2024}. To form an ELM WD with a brown dwarf companion, the main problem is whether there is enough orbital energy to eject the common envelope when the progenitor of the WD has just developed a low-mass helium core. 

To solve this problem, we simulate two stars with masses of $M = 1.0 \rm M_{\sun}$ (metallicity $Z = 0.02$) and $M = 0.8\rm M_{\sun}$ (metallicity $Z = 0.001$) by using the \texttt{Modules for Experiments in Stellar Astrophysics} (\texttt{mesa}, version 10398; \citealt{Paxton2011, Paxton2013, Paxton2015, Paxton2018, Paxton2019}). As the stars evolve, helium cores develop, and the binding energy of the envelope ($E_{\rm {bind}}$) is calculated following \citet{Han1994, Ivanova2013, Han2020}:
\begin{equation}
\label{eq1}
   E_{\rm {bind}} =  - \int_{\rm{core}}^{\rm{surface}}(-\frac{Gm}{r}+U) dm,
\end{equation}
where $G$ is the gravitational constant, $m$ and $r$ are the mass and radius of the star, $U$ is the internal energy \footnote{The choice of boundary for calculating binding energy can influence the final result. Here, we adopt two different boundaries to compute the binding energy: (i) the boundary defined by the helium core (where the hydrogen abundance drops to $X = 0.01$), and (ii) the boundary extending to the helium core mass + 0.01 $\rm M_{\sun}$ \citep{Han1994}. Recently, \citet{Nie2025} showed that a significant fraction of the hydrogen envelope may be retained after common envelope evolution. We tested a boundary extending to the core mass plus 0.03 $\rm M_{\sun}$ for calculating the binding energy.}. We adopt the most commonly used energy formalism to calculate the common envelope ejection process, i.e., the released orbital energy is used to remove the common envelope \citep{Ivanova2013}. The expression is
\begin{equation}
\label{eq2}
    E_{\mathrm{bind}} = \alpha_{\mathrm {CE}}(\frac{Gm_{\mathrm{1,c}}m_{\mathrm{2}}}{2a_{\mathrm{f}}} - \frac{Gm_{\mathrm{1}}m_{2}}{2a_{\mathrm{i}}}),
\end{equation}
where $\alpha_{\rm {CE}}$ is the common envelope ejection efficiency, and $m_{1}$, $m_{\rm {1,c}}$ and $m_{2}$ are the star mass, the helium core mass at the onset of the common envelope phase, and the brown dwarf mass, respectively. Here we adopt the typical brown dwarf mass of 0.06 $\rm M_{\sun}$, and the evaporation effect during the spiral-in process is neglected \citep{Glanz2018}. $a_{\rm i}$ and $a_{\rm f}$ are the initial and final separations, respectively.
During the CE phase, once the brown dwarf fills its Roche lobe, a destructive mass transfer may occur, which leads to a failed common envelope ejection \citep{Nelemans1998}. Therefore, there is a minimum separation, which is given by
\begin{equation}
\label{eq4}
   a_{\mathrm {min,f}} = r_{\mathrm {BD}} \frac{0.6q^{\prime 2/3}+\mathrm{ln}(1+q^{\prime 1/3})}{0.49q^{\prime 2/3}},
\end{equation}
where $r_{\rm {BD}} = 0.1 \rm R_{\sun}$ is adopted for the radius of a brown dwarf \citep{Baraffe2003}, and $q^{\prime} = m_{2}/m_{1,c}$. 

\begin{figure}
  \centering
  \includegraphics[width=0.8\textwidth]{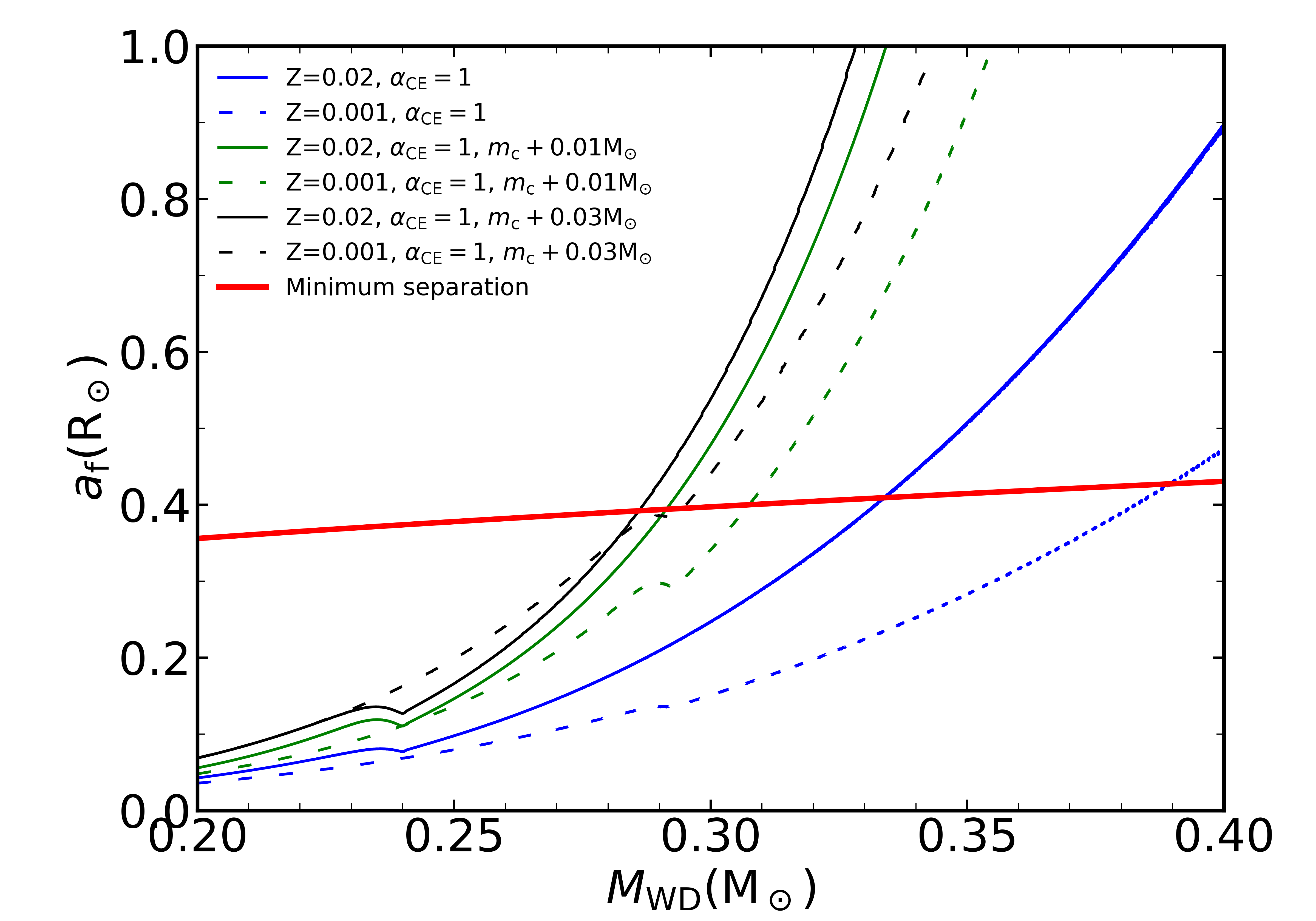}
  \caption{Relationship between ELM WD mass and post-CE binary separation for different models, with $\alpha_{\rm{CE}} = 1$ for all cases. The blue, green, and black curves represent models where the envelope binding energy is calculated using the helium core boundary (defined as where $X = 0.01$), the helium core mass plus 0.01 $M_\odot$, and the helium core mass plus 0.03 $M_\odot$, respectively. For each set, solid and dashed lines correspond to metallicities of $Z = 0.02$ and $Z = 0.001$. The red solid line marks the theoretical minimum binary separation.}
  \label{Fig2}
\end{figure}

The final separation as a function of core mass is presented in Figure ~\ref{Fig2}, where the solid lines are for $M_{1} = 1.0 \rm M_{\sun}, Z = 0.02$, and the dashed lines are for $M_{1} = 0.8 \rm M_{\sun}, Z = 0.001$. In our calculations, some hydrogen envelope remains in the helium core after CE ejection. We assumed that all of the hydrogen envelope is lost after the CE phase to obtain the minimum white dwarf mass. With this assumption, the actual WD mass should be larger than the value shown in Figure \ref{Fig2}.
The red line is the minimum separation. From Figure ~\ref{Fig2}, when assuming all released orbital energy is used for envelope ejection ($\alpha_{\rm {CE}}$ = 1), the minimum white dwarf mass remains above 0.28 $\rm M_{\sun}$, which is larger than the ELM WD masses of KIC 8145411 and HE 0430-2457. 
Therefore, neither of the ELM WDs in KIC 8145411 and HE 0430-2457 can be produced by this process.

\section{SN explosion channel}
\label{sec:3}
We further investigate the SN explosion channel to form wide-orbit ELM WD binaries.
The single ELM WD forms in the inner binary after an SN Ia event, as shown in panel (b) of Figure~\ref{Fig1}. At the beginning, the inner binary contains a massive star (6 - 8 $\rm M_{\sun}$) and a low-mass companion ($\sim$1.8 - 2.2 $\rm M_{\sun}$). The more massive one evolves first, leaving a massive carbon–oxygen (CO) or oxygen-neon (ONe) WD after the first CE phase.
The low-mass star subsequently fills its Roche lobe and transfers mass to the CO WD. During this phase, the CO WD accretes material until reaching the Chandrasekhar limit and ultimately exploding as a Type Ia supernovae \footnote{The WD becomes rapidly rotating during the accretion process, and may not explode immediately upon exceeding the Chandrasekhar limit \citep{Yoon2004, Wang2014}. Here we assume that the CO WD continues to increase its mass and that an explosion occurs at the end of mass transfer.}. 
Finally, a single He WD is left but with a large runaway velocity. The newly formed single ELM WD and the tertiary component form a new binary system. We discuss this channel further in Sections ~\ref{sec:3.1} and ~\ref{sec:3.2}.

\subsection{Binary Evolution Simulations}
\label{sec:3.1}
Using the \texttt{mesa} stellar evolution code, we simulate the formation of a single ELM WD in the SN explosion channel. In this channel, the massive WD (CO/ONe WD) is born first and accretes mass during the subsequent mass transfer phase. Its accretion depends strongly on the mass transfer rate.
The WD can increase its mass only if it accretes at a rate larger than $\dot M_{\rm {stable}}$, where $\dot M_{\rm {stable}}$ is the threshold accretion rate for stable hydrogen burning \citep{Meng2009} and stable helium burning \citep{Kato2004}. In other words, the initial donor mass and orbital separation have a significant effect on the accreted mass of the WD. 
In a recent work, \citet{Li2019} obtained a grid that covers initial CO WDs with masses ranging from 0.45 to 1.1 $\rm M_{\sun}$, and initial donors with masses ranging from 1.0 to 2.2 $\rm M_{\sun}$.
Based on this work, we find that several solutions can satisfy both of the following criteria, i.e., (1) the massive WD can accrete enough mass to reach the Chandrasekhar limit, and (2) the He WD produced from the donor is less than 0.28 $\rm M_{\sun}$.

\begin{figure}
  \centering
  \includegraphics[width=0.8\textwidth]{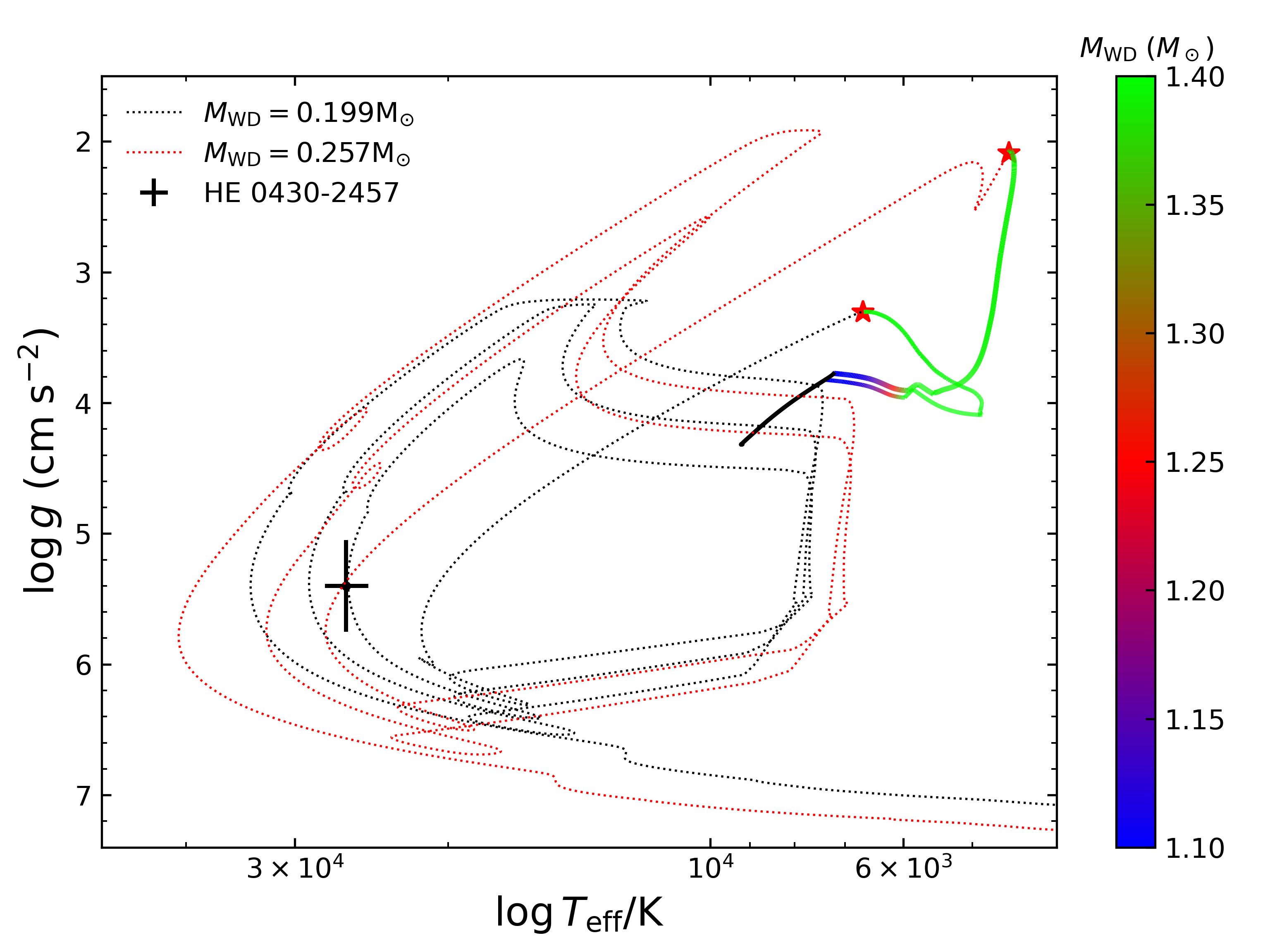}
  \caption{The evolutionary tracks of single ELM WDs from the SN explosion channel in the $T_{\rm {eff}}-{\rm{log}} g$ plane. The colored solid lines show the mass variation of CO WDs. The red stars indicate the end of mass transfer. The plus sign marks the data for HE 0430-2457, taken from \citet{Vos2018b}. 
}
  \label{Fig3}
\end{figure}

Figure ~\ref{Fig3} shows two selected evolutionary tracks that can form single ELM WDs in the $T_{\rm{eff}}-\rm{log} g$ plane.
For the two models, the ELM WDs have an initial progenitor mass of 2.04 $\rm M_{\sun}$, while the CO WDs have an initial mass of 1.1 $\rm M_{\sun}$. The initial orbital periods are 1.14 days and 1.24 days, corresponding to final ELM WD masses of 0.199 $\rm M_{\sun}$ and 0.257 $\rm M_{\sun}$, respectively.
From Figure ~\ref{Fig3}, the progenitor of the ELM WD begins mass transfer at the end of the main sequence (MS), and the CO WD accretes sufficient mass to reach the Chandrasekhar limit before the termination of mass transfer. After mass transfer, the CO WD explodes as an SN Ia, leaving behind a single ELM WD. Due to the thin H-envelope of the ELM WD, it will experience H flashes. By comparing with observations, these evolutionary tracks can reproduce the observed data. This indicates that SN explosions could serve as an evolutionary channel for creating single ELM WDs, which may subsequently form wide-orbit ELM WD binaries.

\subsection{ECCENTRICITY DISTRIBUTION}
\label{sec:3.2}
Both KIC 8145411 and HE 0430-2457 are in nearly circular orbits, and the origin of this phenomenon remains unclear \citep{Vos2018b, Masuda2019}. Then, we explore the eccentricity distributions of ELM WDs in wide orbits after the SN explosion. 

After the explosion of an SN Ia, the companion gains a runaway velocity. Similar to the kick velocity of a neutron star (NS), the runaway velocity of an ELM WD will alter the eccentricity and the binary separation.
We assume that the relative orientation of the inner and outer binary before the SN explosion is random. The runaway velocity is set to the ELM WD's pre-explosion orbital velocity, which is derived from the white dwarf mass-orbital period relation given by \citet{Lin2011}. This velocity may be influenced by the SN explosion; we therefore assume the runaway velocity varies by a factor of 0.8 - 1.2.
To compare with the observations, the ELM WD masses are randomly distributed in the ranges of (0.18 $\rm M_{\sun}$, 0.28 $\rm M_{\sun}$) for HE 0430-2457 and (0.191 $\rm M_{\sun}$, 0.209 $\rm M_{\sun}$) for KIC 8145411, respectively. We use the Monte Carlo method to calculate the final eccentricity and separation after the SN explosion. The initial outer binary separation follows a logarithmic distribution from $20 \rm R_{\sun}$ to $10^{4} \rm R_{\sun}$ \citep{Toonen2020}. The initial outer eccentricity distributions are considered in two cases: (1) $P(e) \approx e^{-0.42}, 0 \leq e < 1 $ \citep{Sana2012}; (2) $e = 0$ for all simulations. The simulations are composed of $2 \times 10^{6}$ binaries. 

\begin{figure}
  \centering
  \includegraphics[width=1.0\textwidth]{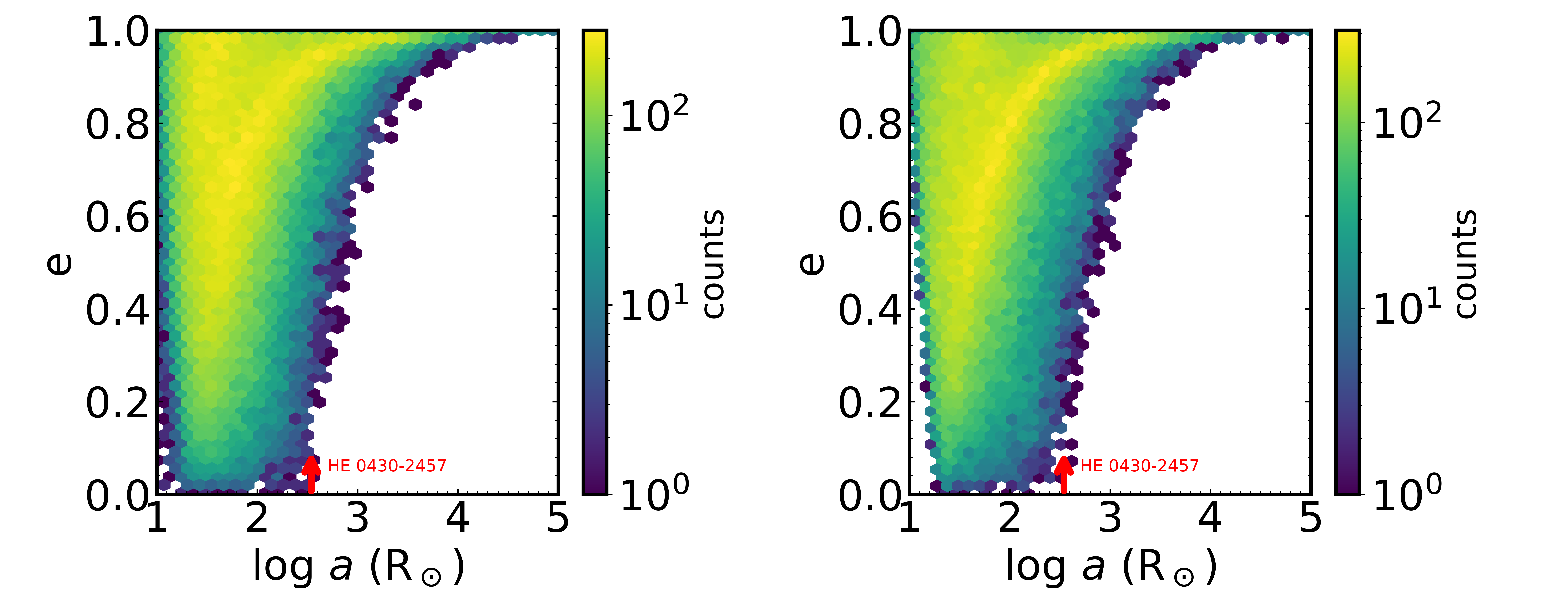}
  \caption{Final separation and eccentricity distributions for HE 0430-2457. Left panel: Results for initially eccentric systems with outer orbit eccentricity distribution $P(e) \approx e^{-0.42}, 0 \leq e < 1 $ case. Right panel: Results for initially circular systems with $e = 0$.}
  \label{Fig4}
\end{figure}

\begin{figure}
  \centering
  \includegraphics[width=1.0\textwidth]{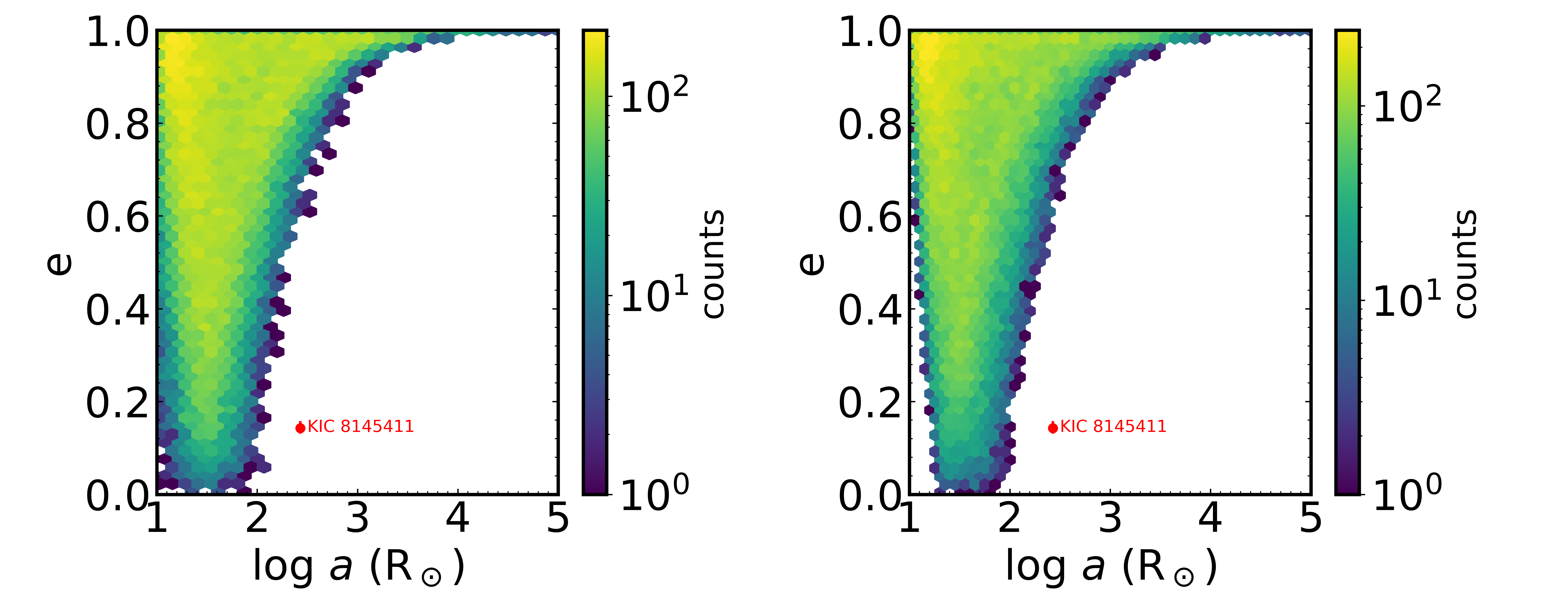}
  \caption{Similar to Figure \ref{Fig4}, but for KIC 8145411.}
  \label{Fig5}
\end{figure}

The distributions of eccentricity and orbital separation are shown in Figures ~\ref{Fig4} and ~\ref{Fig5}, with the left and right panels corresponding to the initial eccentricity distributions for case (1) and case (2), respectively. 
The differences between the two cases are negligible because the runaway velocity dominates over the initial eccentricity in determining the final orbital parameters.
Due to the high runaway velocities of ELM WDs, most binaries become unbound.
For HE 0430-2457, the number of surviving binaries post-SN Ia is 85,294 and 90,070 for cases (1) and (2) respectively, representing $\sim$ 4\% of the total sample. Although the probability is relatively low, the SN explosion channel still has a certain possibility of reproducing the observed parameters of HE 0430-2457.
In contrast, the lower WD mass in KIC 8145411 leads to even higher runaway velocities, resulting in fewer surviving binaries (45,641 for case (1) and 52,587 for case (2)). From Figure ~\ref{Fig5}, our results indicate that this formation channel is unlikely to produce binary systems matching KIC 8145411's observed parameters. Combining this with the analysis in Section~\ref{sec:2}, our conclusion indirectly supports the recent findings of \citet{Yamaguchi2024}, who proposed that KIC 8145411 is in fact a triple system containing a CO WD. Additionally, we predict that this channel is likely to produce wide-orbit ELM WD binaries with $e > 0.5$, which may be detected in the future.

\section{DISCUSSION AND CONCLUSION}

In this work, we investigate the formation of ELM WDs in wide binaries. Two channels are carefully studied: (i) the channel of the merger of an evolved star with a BD and (ii) the SN explosion channel. Our principal conclusions are as follows:

First, in the merger of an evolved star with a BD channel, the progenitor envelope of the WD retains a high binding energy at the onset of the CE phase, which limits production to relatively massive WDs ($> 0.28 \rm M_{\sun}$) and thus cannot reproduce the observations. In comparison, the SN explosion channel can produce ELM WDs ($< 0.28 \rm M_{\sun}$) consistent with the observational data.

Second, through Monte Carlo simulations of the orbital separation and eccentricity distributions between ELM WDs and tertiaries formed via the SN explosion channel, we find that most binaries exhibit high eccentricities. Yet, a small fraction of them are consistent with the observed parameters of HE 0430-2457. However, we are unable to reproduce the observational characteristics of KIC 8145411, which supports the conclusion of \citet{Yamaguchi2024} that this system is a triple system.

%
%


\begin{acknowledgements}
We would like to thank an anonymous referee for very constructive suggestions, which improved the manuscript a lot. We thank Hailiang Chen, Hongwei Ge, Heran Xiong et al. for helpful discussions. The authors gratefully acknowledge the ``PHOENIX Supercomputing Platform'' jointly operated by the Binary Population Synthesis Group and the Stellar Astrophysics Group at Yunnan Observatories, Chinese Academy of Sciences. This work is partially supported by the Natural Science Foundation of China
(Grant no. 11733008, 11521303, 11703081, 11422324), by the National Ten-thousand talents program, by Yunnan province (No. 2017HC018), by Youth Innovation Promotion Association of the Chinese Academy of Sciences (Grant no. 2018076), by the CAS light of West China Program (242300420944) and the Natural Science Foundation of Henan Province (242300420944).
\end{acknowledgements}



\bibliographystyle{raa}
\bibliography{zyy_refs.bib}

\label{lastpage}

\end{document}